# Organizational support for work-family life balance as an antecedent to the well-being of tourism employees in Spain

José Aurelio Medina-Garrido [a,*], José María Biedma-Ferrer [a], Maria Bogren [a,b]

[a] INDESS, Universidad de Cadiz, Spain
[b] Nord University, Norway
[*] Corresponding author (joseaurelio.medina@uca.es)

This is the version submitted for evaluation in the "Journal of Hospitality and Tourism Management" and accepted for publication after the review process.
The final published version can be found at: https://doi.org/10.1016/j.jhtm.2023.08.018
We acknowledge that ELSEVIER holds the copyright of the final version of this work.
Please, cite this paper in this way:
Medina-Garrido, J. A., Biedma-Ferrer, J. M., & Bogren, M. (2023). Organizational support for work-family life balance as an antecedent to the well-being of tourism employees in Spain. Journal of Hospitality and Tourism Management, 57(September), 117–129.

## Abstract

The study of work-family conflict (WFC) and work-family policies (WFP) and their impact on the well-being of employees in the tourism sector is increasingly attracting the attention of researchers. To overcome the adverse effects of WFC, managers should promote WFP, which contribute to increased well-being at work and employees' commitment. This paper aims to analyze the impact of WFP accessibility and organizational support on well-being directly and by mediating the organizational commitment that these policies might encourage. In addition, we also study whether these relationships vary according to gender and employee seniority. To test the hypotheses derived from this objective, we collected 530 valid and completed questionnaires from workers in the tourism sector in Spain, which we analyzed using structural equation modeling based on the PLS-SEM approach. The results show that human resource management must consider the importance of organizational support for workers to make WFP accessible and generate organizational commitment and well-being at work.

# 1 Introduction

Tourism is among the most critical industries globally and has significant economic, social, and environmental impacts. In Spain, tourism is a vital part of the economy, contributing significantly to GDP and employment. However, as the industry continues to grow, the well-being of workers in the sector must also be considered. The tourism industry should prioritize the well-being of its workers by providing adequate support, benefits, and policies that promote work-life balance and a positive work environment. Work-family balance is a critical factor that affects employees' subjective well-being in the hospitality industry. Yang and Jo (2022) found that work-family balance mediated the relationship between recovery experiences and subjective well-being in Chinese hotel employees. Zhao et al. (2020) noted that work-family conflict significantly relates to employees' work, family, and life attitudes in the hospitality industry. Companies need to provide adequate maternity leave policies and practices to improve employee well-being and work-life balance in the hospitality sector. Xu et al. (2021) highlighted the importance of studying maternity leave policies in the US lodging industry and their impact on working mothers' well-being and career advancement. Ariza-Montes et al. (2019) analyzed the working conditions of servers in the European hospitality industry and identified crucial factors that affect psychological well-being or discomfort. As can be seen, the study of WFC and work-family policies (WFP) is increasingly attracting the attention of researchers (Bakar & Salleh, 2015; Li et al., 2015; Zheng & Wu, 2018). The literature shows the negative effect of work on the family and, in turn, that the family negatively affects work (Minnotte et al., 2013, 2015). Rahman et al. (2019) found in a study on the service sector that high job demands, overloaded work, and family responsibilities generate a high level of WFC. These present authors consider it essential to study the impact of WFC on employee well-being.

Addressing WFC is a major concern of those interested in the well-being of employees' work and life (Babin, 2015). In this regard, Kalliath et al. (2017) have pointed out the adverse effects of WFC on employee well-being, and Recuero & Segovia (2021) note that WFC is the main predictor of emotional exhaustion. The literature has shown that WFC reduces well-being and feelings of good health (Winefield et al., 2014). Therefore, a high level of WFC leads to lower job satisfaction, which can lead to a deterioration of physical and psychological health (Amstad et al., 2011; Beauregard & Henry, 2009; Magee et al., 2012). For this reason, the policies implemented by organizations are critical to reducing WFC and its detrimental effects (Aazami et al., 2015).

To overcome WFC, organizational managers need to promote WFP that contribute to increasing well-being at work (Isfianadewi & Noordyani, 2020). Further, there is evidence that WFP positively affects employees' commitment (Mukanzi & Senaji, 2017), and there is a growing number of companies implementing WFP, as they are beneficial for both employees and organizations (Lyonette & Baldauf, 2019; Masuda & Sortheix, 2012; Pérez-Pérez et al., 2017).

Along the same lines as WFP, the literature shows that employees with organizational support have higher levels of emotional well-being (Burke, 2010). Indeed, WFP can be considered a specific type of organizational support. In addition, a supervisor's and colleagues' social support improves employees' well-being and helps reduce work-family conflicts (Lapierre & Allen, 2006).

However, implementing WFP in the organization is a necessary but insufficient condition for overcoming the WFC. In addition, workers must be able to access the WFP without retaliation or other disadvantages (Medina-Garrido, Biedma-Ferrer, & Rodríguez-Cornejo, 2021).

The literature on the impact of WFP on well-being shows several gaps. As noted above, the literature analyzes WFP without differentiating the nature of their existence or their effective accessibility to workers (Medina-Garrido, Biedma-Ferrer, & Ramos-Rodríguez, 2021), and without considering whether they have real organizational support (Burke, 2010; Lapierre & Allen, 2006). In addition, the literature shows evidence of the direct impact of WFP on well-being, but does not study the mediating role that the organizational commitment that these policies can foster might have (Mukanzi & Senaji, 2017). Moreover, previous studies on the impact of WFP and organizational support on organizational commitment and employee well-being highlight the moderating role that seniority (Salami, 2008) and, above all, employee gender (Cukrowska-Torzewska, 2017; Poggesi et al., 2019) can have on these relationships. Therefore, analyzing these relationships should be accompanied by analyzing the moderating role of gender and seniority in the company.

This paper aims to analyze the impact of WFP accessibility and organizational support on well-being directly, and by mediating the organizational commitment that these policies might encourage. In addition, we also study whether these relationships vary according to gender and employee seniority.

To achieve the intended objective, the analysis of the previous literature allowed us to propose several hypotheses relating the accessibility of WFP and organizational support to employees' organizational commitment and well-being. To test these hypotheses, we collected 530 valid and completed questionnaires from workers in the tourism sector, which we analyzed using structural equation modeling based on the PLS-SEM approach.

The results of this research fill the gap in the literature described above and could have important practical implications for management, for which the results will show that human resource management must consider the importance of organizational support for workers to make WFP accessible, and to generate organizational commitment and well-being at work.

This paper is organized as follows. Section two develops the theoretical model and the hypotheses to be tested. Section three defines the sample studied, the study variables, and their measures, and describes the methodology developed. As mentioned above, this section applies structural equation modeling to the data collected. The last sections provide the study's results, the discussion of its implications, and the conclusions.

## 2  Theoretical framework and hypotheses

Work-life balance is a topic of great interest in academia and human resource management, both for its impact on employees' well-being and their productivity and performance in organizations (Wolor et al., 2020). It is intuitive to connect work-life balance with organizational support, employee commitment to the organization, and the well-being these variables generate (Mukanzi & Senaji, 2017; Nordenmark & Alm, 2020; Roemer & Harris, 2018). In this regard, the literature points out that organizational support influences the achievement of the work-life balance (Maszura & Novliadi, 2020), and that improving the work-life balance through WFP increases employees' organizational commitment (Berk & Gundogmus, 2018), and improves workers' well-being (Medina-Garrido, Biedma-Ferrer, & Ramos-Rodríguez, 2021; Nordenmark & Alm, 2020; Roemer & Harris, 2018).

Based on the above, the following is a review of the previous literature on the influence of organizational support on the accessibility of WFPs and the well-being of workers, and the mediating role that organizational commitment may play in these relationships. In addition, the literature shows that it is also relevant to study what could be the moderating role of the variables gender (Medina-Garrido, Biedma-Ferrer, & Ramos-Rodríguez, 2021) and seniority in the company (Salami, 2008).

### 2.1  Accessibility of WFP and well-being

Organizations can benefit from paying attention to the WFC to establish appropriate WFP (Nordenmark & Alm, 2020). Organizations have realized the detrimental effects of WFC on well-being (Sirgy et al., 2020) and employee behavior and, consequently, on the proper functioning of organizational processes (Rahman et al., 2019). In this regard, it has been considered essential for organizations to pay attention to resolving WFC to ensure employee well-being (Neto et al., 2016). The previous literature shows evidence that WFP are positively correlated with both physical well-being and psychological well-being (Medina-Garrido et al., 2017; Semlali & Hassi, 2016)), while the WFC has adverse effects on workers' well-being

(Kalliath et al., 2017) and satisfaction (Deng & Gao, 2017). Given the impact of workers' well-being on their productivity, organizations should implement WFP to improve their well-being (Nordenmark & Alm, 2020). However, implementing WFP is a necessary but insufficient condition for achieving such well-being and adequate job performance. In addition to having WFP in place, these policies need to be easily accessible to workers, and without retaliation at a later stage (Biedma-Ferrer & Medina-Garrido, 2014).

The above arguments allow us to propose the following hypothesis:

> H1. The greater the accessibility of WFP, the greater the well-being.

## 2.2 Organizational support and well-being

Perceived organizational support is defined as employees' belief that the organization values their contributions and pays attention to their well-being (Eisenberger et al., 2016). The literature shows that employees with organizational support have higher levels of emotional well-being (Burke, 2010). The mere perception that such organizational support exists increases the likelihood of achieving higher levels of well-being (Roemer & Harris, 2018). There is evidence that workers with top management contact, decision-making influence, and a well-defined organization experience higher job satisfaction, mental energy, and lower work-related exhaustion (Von Vultée et al., 2007). Along the same line, Lapierre and Allen (2006) highlight that social support from supervisors and peers helps reduce work-family conflict and improves employee well-being. The importance of flexible work arrangements for well-being is noted by Halpern (2005), who found that employees with time-flexible work policies reported less stress. Medina-Garrido et al. (Medina-Garrido, Biedma-Ferrer, & Ramos-Rodríguez, 2021) found that both the existence and accessibility of WFP positively affect emotional well-being. In order to achieve employee well-being and work-life balance, Seiger and Wiese (2009) point out the effectiveness of applying organizational understanding and managerial support for this. Moreover, implementing support systems for improving employee well-being ultimately results in valuable organizational outcomes (Roemer & Harris, 2018).

Based on the above, the following hypothesis can be put forward.

> H2: The greater the organizational support, the greater the well-being.

## 2.3 Organizational support and accessibility of WFP

Employees perceive organizational support positively and significantly affects work-life balance (Maszura & Novliadi, 2020; Yadav & Sharma, 2021). Establishing a supportive organizational environment is a valuable investment for employers (Roemer & Harris, 2018). Organizational support at all levels is essential to facilitate access to WFP. In this context, supervisors, in addition to monitoring the performance of their employees, have an important role to play in reducing WFC (Yadav & Sharma, 2021). More specifically, workers must perceive that both their immediate supervisors and coworkers approve of using WFP and that there are no subsequent reprisals either socially or for their careers (Medina-Garrido et al., 2020). In other words, workers must perceive that the WFP in place are accessible to them without problems.

Considering the above, the following hypothesis can be put forward:

H3: The greater the organizational support, the greater the accessibility to WFP.

## 2.4 Accessibility of WFP and organizational commitment

Organizational commitment is the strength of individuals identifying and engaging with their organization (Ozkan et al., 2020). There is evidence that WFP positively correlate with affective, continuance, and normative commitment (Mukanzi & Senaji, 2017). WFP, with measures such as childcare for employees' children, contribute to employees developing a more significant organizational commitment to their organization (Casper & Harris, 2008). The work-family support that organizations provide to their employees has been considered to make them feel "indebted" to the organization. This feeling encourages employees to make extra efforts in their work or to value the organization they work for more positively. In this regard, WFP contribute to a higher degree of organizational commitment of workers to their organization (Pérez-Pérez et al., 2017; Thompson & Prottas, 2006).

Based on the above, we can state the following hypothesis:

H4. The greater the accessibility of WFP, the greater the commitment to the organization.

## 2.5 Organizational support and organizational commitment

Proper human resource management leads to high employee engagement. To improve organizational commitment, managers must provide appropriate support and ensure that these align with workers' aspirations and needs (Ekowati & Andini, 2008). Employee commitment can improve if organizational support increases (Bibi et al., 2019). The previous literature shows that perceived organizational support positively affects commitment and reduces staff turnover (Winarno et al., 2022; Wu & Liu, 2022). Regarding WFC, the literature has found that employee engagement increases when the organization supports its employees in balancing work and family (Ahmad & Omar, 2010). This positive relationship has already been justified above for the specific case of accessibility to WFP, which are part of organizational support (in itself involving additional factors, both tangible and intangible). Tangible organizational support includes work equipment, monetary resources, benefits, and salaries, among others; and intangible organizational support includes social and emotional support, attention, respect, affiliation, or a motivating and encouraging environment, among others (Addo & Dartey-Baah, 2020; Pahlevan Sharif et al., 2021; Singh et al., 2018). Conversely, it has been studied that if employees feel that the organization does not support them, they will show less commitment toward it (Prottas & Kopelman, 2009). Based on these arguments, it can be stated that:

H5: The greater the organizational support, the greater the organizational commitment.

## 2.6 Organizational commitment and well-being

To ensure employees' well-being at work, organizations must first focus on creating an emotional bond between the employees and the organization (Stevan E. Hobfoll et al., 2018). In turn, organizational commitment can only be ensured when employees perceive well-being at work, not tension (Schaufeli & Bakker, 2004). More specifically, previous work has shown a positive correlation between psychological well-being and organizational commitment (Mebarki et al., 2019; Moser & Galais, 2009; Yalçın et al., 2021). In this sense, organizational commitment decreases employees' exhaustion and increases their well-being at work, improving their commitment to their organization (Chambel & Carvalho, 2022). Moreover, there is evidence that employees' commitment to the organization in which they work not only has a direct effect on their well-being, but also on the well-being of their partners (Rodriguez-Muñoz et al., 2014). The above reasons show that organizational commitment seems to be crucial to ensure employee well-being (Chambel & Carvalho, 2022). Conversely, there is evidence that individuals with low levels of organizational commitment may exhibit negative behaviors related to their lack of well-being at work, such as not showing up for work or being late (Ari & Çağlayan, 2017). The arguments put forward allow us to establish the following hypothesis:

H6: The greater the organizational commitment, the greater the well-being.

Fig. 1 visually shows the relationships established by the hypotheses proposed above.

**Figure 1.** Theoretical model

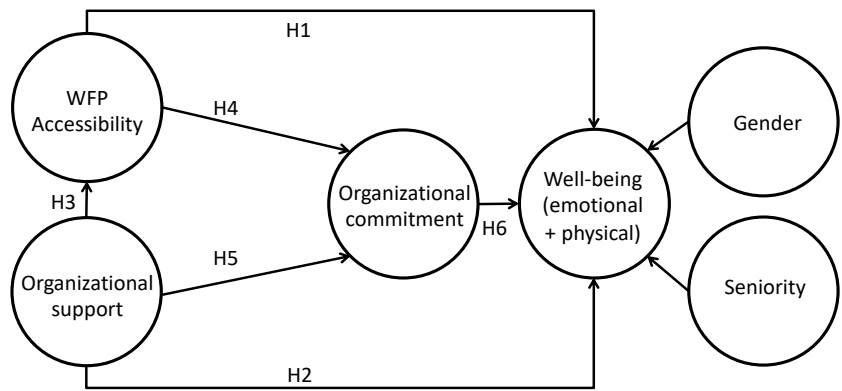

## 2.7 Impact of gender and seniority as moderating variables

Regarding the impact of gender, the literature on WFC shows that women struggle with balancing their multiple roles and responsibilities at work and in their families (Annink & den Dulk, 2012; Cukrowska-Torzewska, 2017; Greenhaus et al., 2003; Poggesi et al., 2019). Compared to men, women in Europe tend to spend, on average, 13 hours a week more on unpaid work than men do (Boye, 2009). The concern for their families and children makes the work-life balance more complex for women (Blanch & Aluja, 2012; Boeckmann et al., 2015) as well as the norm of what a good mother is, concerning who should take care of the home and their children instead of spending time in paid work (Bianchi et al., 2000; Gjerdingen et al., 2000). The results from a study in 25 European countries revealed that in more gender-equal countries (where the labor market is more gender-equal and individuals hold more gender-equal norms), individuals who experience a high level of work-family conflict report lower levels of well-being than individuals who experience high levels of conflict in less gender-equal countries (Hagqvist, 2016; Hagqvist et al., 2017). Hagqvist et al. (Hagqvist et al., 2017) argue that gender context constitutes a critical analytical dimension for understanding the relationship between WFC and well-being, and that it is important for countries to design policies and laws that support women's employment (e.g., quotas for managerial positions, family policies, and antidiscrimination laws), and to include policies supporting a balance between work and family for both men and women.

As for the impact of seniority, several studies from different industries report a positive effect of length of service on organizational commitment (Iqbal et al., 2011; Salami, 2008). A study of two groups of employees within the police force, among police officers and civilian support staff, reveals that the way in which individuals feel they are managed and supported strongly impacts organizational commitment (Dick & Metcalfe, 2001). Another aspect regarding seniority is that people who have worked in an organization for a long time are more likely to have acquired firm-specific skills and are, therefore, more committed to that organization (Gunawan & Marzilli, 2022). Furthermore, Metcalfe and Dick (2000) revealed that the level of organizational commitment varies according to position in the hierarchy, and commitment increases with tenure.

Given the impact that gender and seniority seem to have on the relationships defined in the hypotheses above, we consider their possible role as moderating variables in this study.

# 3 Methodology

## 3.1 Sample and data collection

Initially, 586 questionnaires were collected from workers in the tourism sector in Spain. Applying the pre-assessment of the data proposed by Hair et al. (2017), 56 observations were eliminated for containing a high percentage of missing data. Therefore, the study was conducted with 530 valid questionnaires. Of the 530 effective responses, 228 (43%) were female and 302 (57%) were male respondents, and the average length of service of the sample studied was 18 years (Table 1).

**Table 1.** Descriptive Statistics

|  | Frequency | Percentage | Mean |
|---|---|---|---|
| Gender | 530 |  |  |
| Male | 302 | 57.0 |  |
| Female | 228 | 43.0 |  |
| Years_Worked | 530 |  | 18.1 |

We divided the sample into two subsamples according to the gender of the respondent in order to study the moderating effect of this categorical variable. For this purpose, permutation analysis is the most reliable and recommended test (Chin, 2003). The analysis of the moderating variable Seniority involved studying the interaction effect in each model's relationships and applying the orthogonalization approach (Henseler & Chin, 2010; Little et al., 2006). Considering a statistical power of 0.8 and a default alpha level of 0.05, the sample (n=530) allows the detection of very small effect sizes, which are more difficult to identify, according to Cohen (1988).

## 3.2 Measures

The variables in this study are composites, being design variables or artifacts, where the indicators are linear combinations and compose the variable. Since the indicators of the variables were correlated, we modeled all variables as composites in Mode A, i.e., using correlation weights (Henseler, 2017). Respondents rated the items on a 7-point Likert scale for all variables except Seniority and Gender.

We measured the variable Accessibility with the scales of Anderson, Coffey and Byerly (2002) and Families and Work Institute (2012b, 2012a), totaling 15 items. Some examples of these items are: "There is an unwritten rule in my job that an employee cannot attend to family or personal needs during work hours" (reverse coded), "When I am offered more hours than stipulated in the employment contract, I can refuse to work them without negative consequences at work", or "If I were to apply to use WFP there would be negative consequences regarding advancing my professional career" (reverse coded).

We measured Organizational Support with a 7-item reflective scale from the Families and Work Institute measurement scale (2012b). Examples of the items included are: "Management makes an effort to structure work requirements so that they don't have a negative impact on employees' personal and family lives", or "I have support from coworkers that helps me to manage my work and personal or family lives".

We measured Organizational Commitment using a 16-item scale based on the scales of Boshoff and Mels (2000), Carlson, Kacmar and Williams (2000), and Families and Work Institute (2012b). Some items of this scale are: "The major satisfaction in my life comes from my work", "I would like more time to spend working", or "I am proud to tell others that I am employed by this organization".

As for the variable *Well-Being*, we measured it with 24 indicators associated with the physical and emotional well-being of the individual, based on the scales of Warr (1990) and Kossek, Colquitt and Noe (2001). Examples of some items are: "In relation to the well-being or discomfort you feel at work, during the last few weeks, how many times has your work made you feel Tense, Depressed, Annoying trembling of my hands, or Shortness of breath when not working physically or hard?".

To measure the moderating variables, we used a categorical variable for *Gender* composed of two groups (male and female) and a numerical variable associated with the person's length of service in the company for *Seniority*.

## 3.3 Method

The hypothesis testing followed the methodological recommendations of Hair, Sarstedt, Hopkins and Kuppelwieser (2014) for variables modeled as composites, applying structural equation models based on the PLS-SEM approach. In this sense, the estimates made are consistent. The treatment of missing data consisted of substitution by the mean. The software used to performed the analysis was SmartPLS 3.2.7 (Ringle et al., 2015). The data analysis consisted of the assessment of the overall model, assessment of the measurement model, assessment of the structural model, and finally, the analysis of the moderating effect.

Before proceeding with the data analysis, it is necessary to assess the risk of common method bias. Following the recommendations of Kock (2015) in PLS-SEM, applying the full collinearity test based on variance inflation factors (VIF) is necessary. A VIF value greater than 3.3 indicates pathological collinearity, i.e., the model is contaminated by common method bias. In our study, the maximum value is 1.306 (Table 2), so the model can be considered free of bias.

**Table 2.** Common Bias Method

| Variables | Accessibility | Support | Well-being | Commitment |
|---|---|---|---|---|
| VIF | 1.197 | 1.069 | 1.305 | 1.306 |

# 4 Results

## 4.1 Assessment of the overall model

According to Henseler, Hubona and Ray (2016), the starting point for model evaluation should be the overall goodness of fit of the model, otherwise estimates obtained may be meaningless, and conclusions may be questionable.

The assessment of the overall model involves two stages; the first considers all the model indicators before assessing the measurement model. After assessing the measurement model, the second eliminates those indicators that do not meet the required criteria.

The overall model study used the standardized root mean square residual (SRMR) as an approximate fit measure (Henseler, Hubona, et al., 2016). The SRMR threshold is 0.08 (Hu & Bentler, 1998). The results yielded values of 0.079 before removing the indicators and 0.058 ex-post. According to these data, the model is better by removing the indicators that present reliability and validity problems. Thus, we have an approximately true model.

## 4.2 Assessment of the measurement model

The model of this research contains composites estimated in Mode A. The first step requires analyzing reliability (individual and construct reliability), i.e., that the indicators actually measure what they are intended to measure. Subsequently, we need to check validity (convergent and discriminant validity), i.e., that the measurement is stable and consistent.

Analyzing the reliability of individual items requires an examination of the loadings of the indicators with their respective constructs. According to Carmines & Zeller (1979), to accept the indicators, they must have loads greater than 0.707. However, some researchers indicate that indicators with loads between 0.4 and 0.707 should not be eliminated if they do not pose problems for the rest of the stages of the measurement model. In our study, we have eliminated three indicators corresponding to the Accessibility variable and five indicators of the Organizational Commitment variable. These eliminated indicators had very low loads (below 0.4). In addition, we removed three more indicators of the Accessibility variable, one of the Organizational Support variable, and three of the Organizational Commitment variable, to meet the requirement of convergent validity an thus increasing the value of the average variance extracted (AVE).

Construct reliability determines whether items measuring a construct are similar in their scores. To analyze construct reliability, we used measures corresponding to composite reliability (Werts et al., 1974), the reliability construct of Dijkstra & Henseler (2015), and suggest values (above 0.8) of Nunnally and Bernstein (1994) .

Subsequently, we studied convergent validity with the average variance extracted (AVE) to verify that the indicators represent a single underlying construct. Acceptable AVE values are those above 0.5 (Fornell & Larcker, 1981).

Discriminant validity assesses the degree to which a construct differs from other constructs. The HTMT ratio developed by Henseler, Ringle, and Sarstedt (2016) and the Fornell and Larcker (1981) Criterion, which uses the matrix of correlations between variables, are two ways to assess this discriminant validity.

Once we have eliminated the indicators described above, the requirements of construct reliability, convergent validity, and discriminant validity are satisfied. Tables 3, 4, and 5 show the results of these analyses.

**Table 3.** Measurement model

| Construct/Indicators | Loads | α Cronbach | CR | ρA | AVE |
|---|---|---|---|---|---|
| **ACCESSIBILITY (Composite, Mode A)** | | 0.898 | 0.917 | 0.904 | 0.553 |
| ACC_Attitude | 0.778 | | | | |
| ACC_Advancing | 0.840 | | | | |
| ACC_Committed | 0.759 | | | | |
| ACC_Choice | 0.747 | | | | |
| ACC_Income | 0.709 | | | | |
| ACC_Needs | 0.758 | | | | |

| | | CR | ρA | | AVE |
|---|---|---|---|---|---|
| ACC_Permits | 0.787 | | | | |
| ACC_Rule | 0.594 | | | | |
| ACC_Supervisor | 0.694 | | | | |
| **SUPPORT (Composite, Mode A)** | | 0.828 | 0.875 | 0.830 | 0.539 |
| APO_Performance | 0.670 | | | | |
| APO_Requirements | 0.780 | | | | |
| APO_Impact | 0.724 | | | | |
| APO_Reporting | 0.677 | | | | |
| APO_Fair | 0.786 | | | | |
| APO_Needs | 0.761 | | | | |
| **WELL-BEING (Composite, Mode A)** | | 0.960 | 0.962 | 0.967 | 0.516 |
| BI_Happy | 0.774 | | | | |
| BI_Calm | 0.708 | | | | |
| BI_Contented | 0.730 | | | | |
| BI_Depressed | 0.802 | | | | |
| BI_Unhappy | 0.780 | | | | |
| BI_Angry | 0.752 | | | | |
| BI_Enthusiastic | 0.663 | | | | |
| BI_Unsettled | 0.760 | | | | |
| BI_Irritated | 0.806 | | | | |
| BI_Angry | 0.766 | | | | |
| BI_Optimistic | 0.721 | | | | |
| BI_Concerned | 0.732 | | | | |
| BI_Relaxed | 0.689 | | | | |
| BI_Tensed | 0.744 | | | | |
| BI_Sad | 0.815 | | | | |
| BI_Air | 0.694 | | | | |
| BI_Appetite | 0.622 | | | | |
| BI_Sleeping | 0.712 | | | | |
| BI_Stomach | 0.698 | | | | |
| BI_Strong | 0.695 | | | | |
| BI_Dizziness | 0.612 | | | | |
| BI_Rapid | 0.694 | | | | |
| BI_Sweating | 0.630 | | | | |
| BI_Tremors | 0.587 | | | | |
| **COMMITMENT (Composite, Mode A)** | | 0.860 | 0.889 | 0.908 | 0.513 |
| COM_Things | 0.502 | | | | |
| COM_Desiring | 0.610 | | | | |
| COM_Destination | 0.540 | | | | |
| COM_Chosen | 0.850 | | | | |
| COM_Proud | 0.865 | | | | |
| COM_Opinion | 0.853 | | | | |
| COM_Satisfactions | 0.518 | | | | |
| COM_Site | 0.848 | | | | |

Note: CR: Composite reliability. ρA: Dijkstra-Henseler. AVE: Average Variance Extracted.

**Table 4.** Discriminant validity. Fornell and Larcker

|  | Accessibility | Support | Well-being | Commitment |
|---|---|---|---|---|
| Accessibility | 0.743 | | | |
| Support | 0.398 | 0.734 | | |
| Well-being | 0.345 | 0.389 | 0.719 | |
| Commitment | 0.263 | 0.476 | 0.451 | 0.716 |

Note: For discriminant validity to be verified, each element of the diagonal must be greater than the rest of the same row or column.

**Table 5.** Discriminant validity. HTMT

|  | Accessibility | Support | Well-being | Commitment |
|---|---|---|---|---|
| Accessibility | | | | |
| Support | 0.449 | | | |
| Well-being | 0.356 | 0.412 | | |
| Commitment | 0.272 | 0.540 | 0.413 | |

Note: In order for discriminant validity to be verified, the elements of the table (square root of the average variance extracted (AVE)) must be less than 0.8 or 0.9, as the case may be.

### 4.3 Structural model

After confirming that the model is reliable and valid, we analyzed the hypotheses in the model. We used the bootstrapping technique for resampling with 5,000 samples to assess the magnitude, sign, and significance of the relationships between the variables. We also analyzed the model's predictive power utilizing the coefficient of determination ($R^2$) of the endogenous variables, and analyzed the decomposition of the variance explained to determine the importance of each variable in the dependent variable. Finally, we assessed the size of the effects following Cohen's (1988) rules. Table 6 shows these results.

**Table 6.** Structural model. Direct effects

|  | Direct effect | p-value | t-value | CI | Supported | Explained variance | $f^2$ |
|---|---|---|---|---|---|---|---|
| **Well-being ($R^2$ = 0.275)** | | | | | | | |
| H1: Accessibility | 0.198 | 0.000 | 4.236 | (0.106;0.287) | Yes | 6.83% | 0.045 |
| H2: Support | 0.155 | 0.001 | 3.188 | (0.059;0.249) | Yes | 6.03% | 0.023 |
| H6: Commitment | 0.325 | 0.000 | 7.546 | (0.240;0.410) | Yes | 14.66% | 0.112 |
| **Commitment ($R^2$ = 0.234)** | | | | | | | |
| H4: Accessibility | 0.088 | 0.063 | 1.860 | (-0.003;0.180) | No | 2.31% | 0.008 |
| H5: Support | 0.442 | 0.000 | 9.730 | (0.356;0.531) | Yes | 21.04% | 0.214 |
| **Accessibility ($R^2$ = 0.158)** | | | | | | | |
| H3: Support | 0.398 | 0.000 | 9.315 | (0.316;0.484) | Yes | 15.84% | 0.188 |

Note: Bootstrapping process based on n=5,000 samples. Hypotheses using a two-tailed Student's t-distribution. (CI 95%).

The R² values obtained in this model show a low predictive power for the Accessibility variable and a moderate predictive power for the Well-being and Organizational Commitment variables (Chin, 1998). All hypotheses are supported except H4, so that WFP Accessibility does not affect Organizational Commitment at work. On the other hand, the rest of the hypotheses are fulfilled with a small effect for H1, H2, and H6, and a moderate effect for H3 and H5, according to Cohen's (1988) tables.

## 4.4 Moderating variables

After analyzing the hypotheses that form the theoretical model, we studied the possible moderating role in the model of the variables Seniority and Gender. These moderating effects were analyzed differently since the Seniority variable is continuous, while the Gender variable is categorical.

To evaluate the interaction effect of Seniority on the relationships proposed in the model, we compared the model with and without this interaction effect. Following Little et al. (2006), we applied the orthogonalization approach to test the moderating effect on each of the relationships included in our model. Table 7 shows that Seniority in the company does not serve as an antecedent for any of the endogenous variables, but it does moderate the existing relationship between Organizational Support and Commitment. The stronger the moderating variable Seniority, the lower the relationship's strength.

**Table 7.** Structural model with continuous moderation of the Seniority Variable

|  | Direct effect | p-value | t-value | CI | Supported |
|---|---|---|---|---|---|
| **Well-being (R² = 0.275)** | | | | | |
| Seniority | 0.011 | 0.760 | 0.305 | (-0.057;0.079) | No |
| Accessibility | -0.014 | 0.855 | 0.183 | (-0.133;0.147) | No |
| Support | -0.065 | 0.461 | 0.737 | (-0.172;0.154) | No |
| Commitment | -0.058 | 0.574 | 0.562 | (-0.162;0.182) | No |
| **Commitment (R² = 0.234)** | | | | | |
| Seniority | 0.038 | 0.299 | 1.039 | (-0.033;0.110) | No |
| Accessibility | -0.056 | 0.369 | 0.899 | (-0.141;0.123) | No |
| Support | -0.089 | 0.041 | 2.046 | (-0.184;-0.032) | Yes |
| **Accessibility (R² = 0.158)** | | | | | |
| Seniority | -0.016 | 0.692 | 0.396 | (-0.094;0.061) | No |
| Support | 0.048 | 0.530 | 0.628 | (-0.122;0.152) | No |

Note: Bootstrapping process based on n=5,000 subsamples. Calculated two-tailed Student's t-test hypotheses. (CI 95%).

As for the moderating variable Gender, we performed a multigroup analysis to test the interaction effect. For this purpose, we divided the sample into two subsamples. Comparing the models for men and women shows some differences, but we do not know if they are significant. Therefore, we applied a multigroup analysis based on permutations (Chin, 2003). We applied the previous analysis of the measurement invariance issue to ensure that the differences between groups are due to the path coefficients and not to the parameters of the measurement model. When performing this preliminary analysis, we found total invariance in one case and partial invariance in the rest (see Table 8). We then applied the permutation analysis developed by Chin (2003) to evaluate whether there were significant differences between each pair of groups.

**Table 8.** Question of measurement invariance (MICOM)

| Groups/Construct | Step 1 Configuration invariance | Step 2 Composite invariance | | | Step 3a Equality of variances | | | | Step 3b Equality of means | | | | Invariance of the total supported measurement |
|---|---|---|---|---|---|---|---|---|---|---|---|---|---|
| | | Original correlation | 5% | Invariance of the supported partial measurement | Difference between original variances | 2.5% | 97.5% | Equal | Original mean difference | 2.5% | 97.5% | Equal | |
| **Male - Female** | | | | | | | | | | | | | |
| Accessibility | Yes | 0.997 | 0.994 | Yes | 0.176 | -0.239 | 0.264 | Yes | 0.321 | -0.173 | 0.193 | No | No |
| Support | Yes | 0.998 | 0.995 | Yes | -0.146 | -0.250 | 0.240 | Yes | 0.093 | -0.179 | 0.181 | Yes | Yes |
| Well-being | Yes | 0.998 | 0.994 | Yes | -0.093 | -0.204 | 0.209 | Yes | 0.184 | -0.166 | 0.177 | No | No |
| Commitment | Yes | 0.995 | 0.990 | Yes | -0.270 | -0.204 | 0.213 | No | -0.046 | -0.185 | 0.166 | Yes | No |

Subsequently, we performed the permutation analysis in all cases since we had invariance of the measure. We found that when gender moderates, there are only significant differences in the relationship between accessibility of WFP and well-being (see Table 9).

**Table 9.** Multi-group analysis based on the permutation test

| Groups/Direct effects | Group 1 | | | Group | | | Permutation | Significance |
|---|---|---|---|---|---|---|---|---|
| | R² | Direct effect | p-value | R² | Direct effect | p-value | p-value | |
| **Male - Female** | | | | | | | | |
| *Well-being* | 0.282 | | | 0.299 | | | | |
| Accessibility | | 0.301 | 0.000 | | 0.017 | 0.820 | 0.003 | Yes |
| Support | | 0.101 | 0.104 | | 0.237 | 0.001 | 0.190 | No |
| Commitment | | 0.304 | 0.000 | | 0.371 | 0.000 | 0.444 | No |
| *Commitment* | 0.188 | | | 0.311 | | | | |
| Accessibility | | 0.078 | 0.181 | | 0.136 | 0.081 | 0.567 | No |
| Support | | 0.400 | 0.000 | | 0.481 | 0.000 | 0.381 | No |
| *Accessibility* | 0.125 | | | 0.216 | | | | |
| Support | | 0.353 | 0.000 | | 0.465 | 0.000 | 0.207 | No |

## 5 Discussion

The results of this study show that organizational support for employees is vital for achieving organizational commitment and well-being at work, both emotionally and physically. These results confirm the findings of previous studies, which have already highlighted the relationship between organizational support and organizational commitment (Bibi et al., 2019; Winarno et al., 2022; Wu & Liu, 2022) on the one hand, and the impact of organizational support on well-being at work (Burke, 2010; Roemer & Harris, 2018) on the other.

This organizational support also positively affects the WFP implemented in the organization, as stated by Maszura and Novliadi (2020) and Yadav and Sharma (2021). The results of our study go beyond the existing literature and show that organizational support not only favors the existence of WFP, but also promotes accessibility of these policies for workers without retaliation or detriment.

The accessibility of these WFP, in turn, positively impacts well-being at work. Previous work has shown that the existence of WFP positively correlates with physical and psychological well-being (Semlali & Hassi, 2016), and WFP reduce the impact of WFC, which has adverse effects on well-being (Kalliath et al., 2017). In contrast to previous empirical work, our study assumes that the mere existence of WFP may be insufficient therefore they must also be accessible. However, there is scarce literature that empirically analyses the accessibility of WFPs (Medina-Garrido et al., 2020; Medina-Garrido, Biedma-Ferrer, & Ramos-Rodríguez, 2021). According to this literature, workers may have difficulties accessing (or not) WFP for various reasons, such as fear of retaliation, a lack of commitment to the organization, or harming their professional careers. Considering this gap in the literature, our work goes further and investigates whether WFPs are really accessible to workers, and their impact on physical and emotional well-being. This consideration of the accessibility of WFPs constitutes a value added to academic empirical research.

Therefore, according to the results of our study and consistent with the previous literature, well-being at work depends on WFP being accessible (Medina-Garrido et al., 2017) and effectively resolving problems with WFC (Nordenmark & Alm, 2020). Additionally, well-being also depends on organizational support (Roemer & Harris, 2018) and the organizational commitment that this support generates in workers (Chambel & Carvalho, 2022; Stevan E. Hobfoll et al., 2018).

The analysis of the possible impact of the moderating variables *Gender* and *Seniority* shows that they do not significantly impact the model. However, the results reveal some moderating effects on a couple of relationships. More specifically, *Gender* has a moderating effect on the relationship between WFP accessibility and the well-being that these policies generate. This moderating effect could be consistent with the female-dominated nature of the WFC (Poggesi et al., 2019). According to the previous literature, WFP may be less accessible for women than for men due to the wage penalties women face, for example, when returning from parental leave (Cukrowska-Torzewska, 2017).

Concerning the moderating effect of seniority, Salami (2008) already found evidence that job tenure was a significant predictor of organizational commitment. Consistent with this work, in our study, seniority also shows a moderating role in the impact of organizational support on organizational commitment.

Regarding the implications of this study for the literature, this paper addresses some gaps in the impact of WFP on well-being. As discussed, the results of this study provide empirical evidence of the impact of the accessibility of WFP on well-being. In contrast, to our knowledge, the previous studies have focused only on the existence of WFP without discerning whether such policies are accessible to workers without problems and subsequent reprisals. In addition, this work also provides original evidence that supportive organizations show higher levels of accessibility to WFP. Another contribution of this study is in demonstrating the mediating role of organizational commitment between organizational support and employee well-being.

Regarding the implications for managerial practice, the results of this research could have important practical implications for human resource management in the tourism sector. In this regard, the manager must consider the importance of organizational support for workers to make WFP accessible and generate organizational commitment and well-being at work.

# 6 Conclusion

The study of work-family conflict (WFC) and work-family policies (WFP) and their impact on the well-being of employees in the tourism sector is increasingly attracting the attention of researchers. Promoting accessible WFP increases well-being at work (Isfianadewi & Noordyani, 2020) and performance (Medina-Garrido et al., 2017; Medina-Garrido, Biedma-Ferrer, & Ramos-Rodríguez, 2021). Furthermore, there is evidence that emotional well-being increases when there are employees' commitment (Mebarki et al., 2019; Yalçın et al., 2021) and organizational support (Burke, 2010).

This paper aimed to analyze the impact of the accessibility and organizational support of WFP on well-being, both directly and through organizational commitment. We also studied whether the gender and seniority of employees played a moderating role in these relationships. To test the proposed hypotheses, we collected 530 valid questionnaires from workers in the tourism sector, which we analyzed using structural equation modeling based on the PLS-SEM approach.

This research showed the importance of organizational support for workers to make WFP accessible and to generate organizational commitment and well-being at work. However, the impact of organizational support on organizational commitment varies according to the employee's seniority. In addition, the accessibility of WFP also has a direct and positive effect on well-being. Nevertheless, depending on the gender of the employee this relationship can be significantly different.

These results provide value added to the literature on WFP and well-being. This work provides evidence that supportive organizations show higher accessibility to WFP. Findings also demonstrate that organizational commitment mediates organizational support and employee well-being. Moreover, this evidence provides practical implications for management, which must promote organizational support and make WFP accessible to workers to generate organizational commitment and well-being at work.

This study has some limitations. As noted, gender moderates the relationship between WFP accessibility and the well-being that these policies generate. However, the analysis does not allow us to discern whether moderation implies greater or lesser well-being for women or men, given equal accessibility to WFP. Similarly, seniority in the organization also plays a moderating role in the impact of organizational support on organizational commitment. As for gender, it would be necessary to study in greater depth whether greater seniority implies a greater or lesser commitment to equal organizational support. Another limitation of this study is generated by the cultural and regulatory context of the country where the study was conducted, Spain. An interesting line of work for future studies to overcome this limitation would be replicating the current fieldwork in different countries. This replication would allow us to make international comparisons. It would be equally interesting to repeat this study at different points in time to make longitudinal comparisons. These two lines of work would make the results of this study more generalizable to the general population. Finally, other interesting future lines of research would be to study the impact of WFP, organizational support, and organizational commitment on other variables, such as performance, work stress, turnover, absenteeism, and satisfaction, among others.